\documentclass[mathpazo]{cicp}

\newcommand{\url}[1]{{\tt#1}}

\begin{document}
\title{Event-by-event simulation of the Hanbury Brown-Twiss experiment with coherent light}



\author[F. Jin et.~al.]{F. Jin\affil{1}, H. De Raedt\affil{1}\comma\corrauth, and K. Michielsen\affil{2}}
\address{\affilnum{1}\ Department of Applied Physics, Zernike Institute for Advanced Materials,
University of Groningen, Nijenborgh 4, NL-9747 AG Groningen, The Netherlands \\
\affilnum{2}
Institute for Advanced Simulation, J\"ulich Supercomputing Centre, Research Centre Juelich, D-52425 Juelich, Germany
}
\email{{\tt F.Jin@rug.nl} (F. Jin), {\tt h.a.de.raedt@rug.nl} (H. De Raedt), {k.michielsen@fz-juelich.de} (K. Michielsen)}


\begin{abstract}
We present a computer simulation model for the Hanbury Brown-Twiss experiment that is entirely particle-based
and reproduces the results of wave theory.
The model is solely based on experimental facts, 
satisfies Einstein's criterion of local causality and does not require knowledge 
of the solution of a wave equation.
The simulation model is fully consistent with earlier work
and provides another demonstration that it is possible to give a 
particle-only description of wave phenomena, rendering the concept of wave-particle duality superfluous.
\end{abstract}

\pacs{02.70.-c 
,
42.50.Ar
,
42.50.Dv
,
03.65.-w
}
\keywords{Computational techniques, Hanbury Brown-Twiss effect, light coherence, light interference, quantum theory}

\maketitle

\section{Introduction}

Computer simulation is widely regarded as complementary to theory and experiment~\cite{LAND00}.
Usually, the fundamental theories of physics provide the framework to
formulate a mathematical model of the observed phenomenon, 
often in terms of differential equations.
Solving these equations analytically is a task that is often prohibitive
but usually it is possible to study the model by computer simulation.
Experience has shown that computer simulation is a very powerful approach
to study a wide variety of physical phenomena.
However, recent advances in nanotechnology are paving the way to 
prepare, manipulate, couple and measure single microscopic systems
and the interpretation of the results of such experiments 
requires a theory that allows us to construct processes 
that describe the individual events that are being observed.
Such a theory does not yet exist.
Indeed, although quantum theory (QT) provides a recipe to compute the frequencies for observing events,
it does not describe individual events, such as
the arrival of a single electron at a particular position 
on the detection screen~\cite{FEYN65,HOME97,TONO98,BALL03}.
Thus, we face the situation that we cannot rely on an established physical theory 
to build a simulation model for the individual processes that we observe in real experiments.
Of course, we could simply use pseudo-random numbers to generate events according to the probability distribution
that is obtained by solving the Schr{\"o}dinger equation.
However, that is not what the statement ``QT does not describe individual events'' means.
What it means is that QT tells us nothing about the underlying processes that
give rise to the frequencies of events observed after many of these events have been recorded.
Therefore, in order to gain a deeper understanding in the processes that
cause the observed event-based phenomena, it is necessary to model
these processes on the level of individual events without using QT. 
The challenge therefore is to find algorithms that simulate, event-by-event,
the experimental observations that, for instance, interference patterns appear only after a large
number of individual events have been recorded by the detector~\cite{GRAN86,TONO98}, without
first solving the Schr\"odinger equation.

In this paper, we leave the conventional line-of-thought,
postulating that it is fundamentally impossible to give a logically consistent 
description of the experimental results in terms of causal processes of individual events.
In other words, we reject the dogma that there is no explanation that goes 
beyond the quantum theoretical description in terms of averages over many events
and search for an explanation of the experimental facts in terms of elementary,
particle-like processes.
It is not uncommon to find in the recent literature,
statements that it is impossible to simulate quantum phenomena by classical processes.
Such statements are thought to be a direct consequence of Bell's theorem~\cite{BELL93} but
are in conflict with other work that has pointed out the irrelevance of Bell's theorem
~\cite{PENA72,FINE74,FINE82,FINE82a,FINE82b,MUYN86,JAYN89,BROD93,FINE96,KHRE99,SICA99,BAER99,%
HESS01,HESS05,ACCA05,KRAC05,SANT05,MORG06,KHRE07,ADEN07,NIEU09,MATZ09,RAED09a,KARL09}.
This conclusion is supported by several explicit examples that prove
that it is possible to construct algorithms that satisfy
Einstein's criterion for locality and causality, yet reproduce
{\sl exactly} the two-particle correlations of a quantum system in the singlet state,
without invoking any concept of QT~\cite{RAED06c,RAED07a,RAED07b,RAED07c,RAED07d,ZHAO08}.
It is therefore an established fact
that purely classical processes can produce the correlations
that are characteristic for a quantum system in an entangled state,
proving that from the viewpoint of simulating quantum phenomena on a digital computer,
Bell's no-go theorem is of no relevance whatsoever.

This present paper builds on earlier
work~\cite{RAED05d,RAED05b,RAED05c,MICH05,RAED06c,RAED07a,RAED07b,RAED07c,ZHAO07b,ZHAO08,ZHAO08b,JIN09a,JIN09c}
that demonstrates that quantum phenomena can be simulated on the level of individual events
without first solving a wave equation and even invoking concepts of QT, wave theory or probability theory.
Specifically, we have demonstrated that it is possible to construct event-by-event proceses,
that reproduce the results of QT for 
single-photon beam-splitter and Mach-Zehnder interferometer experiments~\cite{GRAN86},
Einstein-Podolsky-Rosen-Bohm experiments with photons~\cite{ASPE82a,ASPE82b,WEIH98},
Wheeler's delayed-choice experiment with single photons~\cite{JACQ07},
quantum eraser experiments with photons~\cite{SCHW99},
double-slit and two-beam single-photon interference,
quantum cryptography protocols,
and universal quantum computation~\cite{MICH05,RAED05c}.
According to the theory of quantum computation, the latter proves that at least in principle,
we can construct particle-like, event-by-event processes that can simulate any quantum system~\cite{NIEL00}.
Some interactive demonstration programs are available for download~\cite{COMPPHYS,MZI08,DS08}.

In this paper, we extend the range of applications of the event-based simulation approach
by demonstrating that the event-based algorithms, used in our previous work,
can be re-used, without modification, to build a particle-only simulation model
for another fundamental physics experiment, the Hanbury Brown-Twiss (HBT) experiment~\cite{HBT56a}.
The HBT effect refers to a variety of correlation and anti-correlation effects in the intensities received 
by two or more detectors from a beam of particles~\cite{GLAU63a,GLAU63b,MAND99}. 
According to common lore, when a HBT experiment is performed using single-particle detectors, 
the HBT effect is attributed to the wave-particle duality of the beam.
In this paper, we present a particle-only model of the HBT effect,
demonstrating that it is possible to construct causal, particle-like
processes that describe the experimental facts without invoking concepts of QT.
     
\begin{figure}[t]
\begin{center}
\includegraphics[width=12cm]{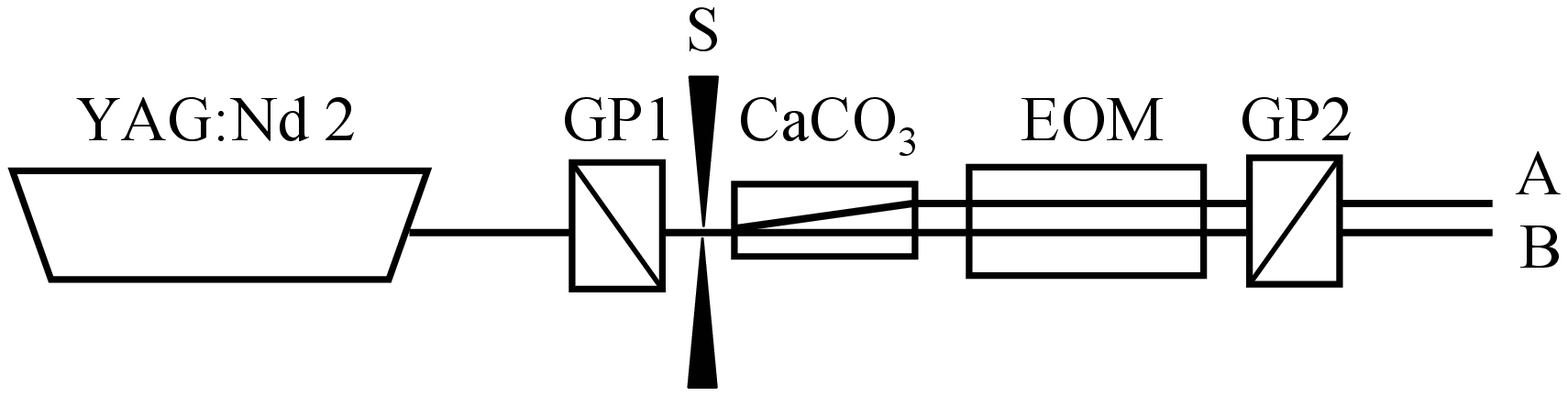}
\includegraphics[width=12cm]{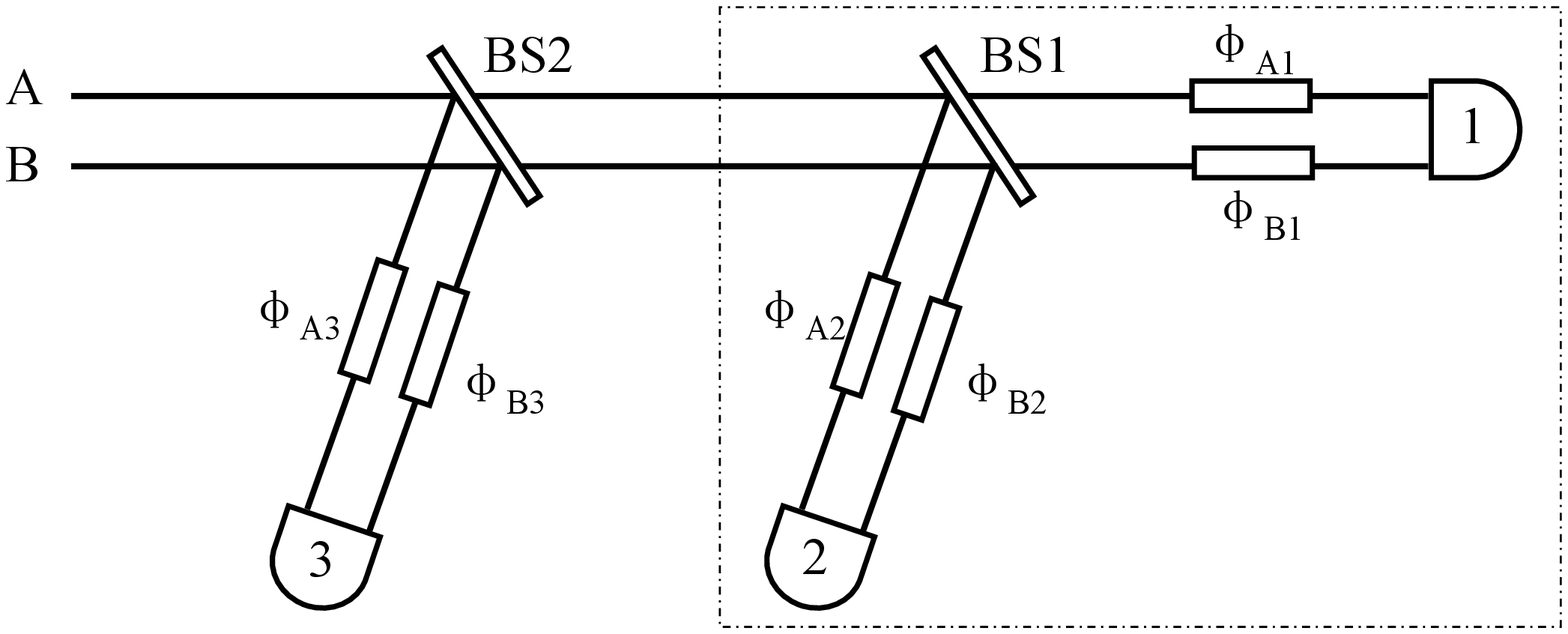}
\caption{Schematic picture of a HBT experiment~\cite{AGAF08}.
Top: Source. Coherent light, generated by a YAG laser, is sent 
through the Gan prism GP1,a single slit S, a beam splitter (a CaCO$_3$ crystal), an electro-optic modulator (EOM)
and another Gan prism GP2 to produce two beams $A$ and $B$ as
if they would have emerged from a double slit separated by 1.3 mm~~\cite{AGAF08}.
The EOM is switched rapidly to destroy the first-order coherence between beams $A$ and $B$. 
Bottom: The interferometer consists of two beam splitters $BS1$ and $BS2$ and phase shifters
$\phi_{An}$ and $\phi_{Bn}$ ($n=1,2,3$). 
Light intensity is measured by the three detectors $D_1$, $D_2$ and $D_3$.
}
\label{hbt2}
\end{center}
\end{figure}

As a concrete realization, we consider a recent HBT experiment~\cite{AGAF08},
a schematic picture of which is shown in Fig.~\ref{hbt2}.
A radiation source, a frequency doubled $Q$-switched Nd:YAG laser with wavelength $532 nm$, is used. 
The coherent light from this source is split by a beam splitter.
The electro-optical modulator (EOM) erases the first-order interference
of the light~\cite{AGAF08}.
The two beams that emerge are labeled $A$ and $B$, see Fig.~\ref{hbt2}(top).
Then, the two beams are sent to the three detectors through two beam splitters (BS), see Fig.~\ref{hbt2}(bottom).
After measuring the coincidences of three detectors by means of a triple coincidence circuit (TCC), 
the third-order intensity interference pattern is observed~\cite{AGAF08}.

The purpose of this paper is to demonstrate that one can construct a simulation model
of this experiment that
\begin{itemize}
\item{is a one-to-one copy of the experimental setup such that
each device in the real experiment has a counterpart in the simulation algorithm
}
\item{is event-based and satisfies elementary physical (Einstein's) requirements of local causility}
\item{reproduces the results of wave theory by means of particles only.}
\end{itemize}

The structure of the paper is as follows.
In Section~2, we briefly review the wave theory of second and third-order coherence.
The simulation model is described in Section~3.
Section~4 presents our simulation results and a discussion thereof.
Our conclusions are given in Section~5.

\section{Wave theory}

\begin{figure}[t]
\begin{center}
\includegraphics[width=12cm]{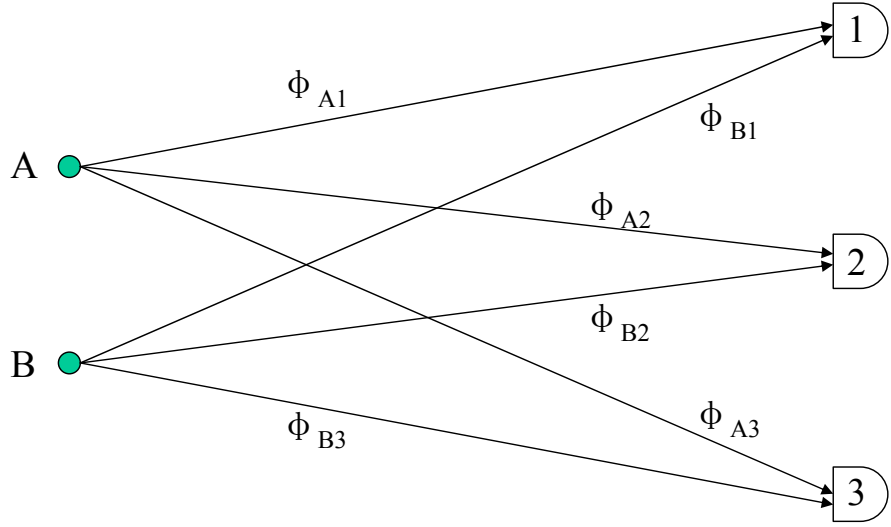}
\caption{Schematic picture of third order intensity correlation.
Photons emitted from sources $A$ and $B$ are registered by three detectors $D_1$, $D_2$ and $D_3$.
$\phi_{An}$ and $\phi_{Bn}$ ($n=1,2,3$) are the phases accumulated during their flight 
from sources $A$ or $B$ to the $n$-th detector.
}
\label{hbt}
\end{center}
\end{figure}

Conceptually, the experiment of Fig.~\ref{hbt2} can be viewed as 
a double-slit type experiment with three detectors, as shown in Fig.~\ref{hbt}.
Assume that source A emits coherent light with amplitude $\alpha$
and that source B emits coherent light with amplitude $\beta$.
Thus, according to the superposition principle, the total amplitude falling on the $n$-th detector ($n=1,2,3$) is
\begin{equation}
a_n =\alpha e^{i \phi_{An}} + \beta e^{i \phi_{Bn}},
\end{equation}
where $\phi_{An}$ ($\phi_{Bn}$) is the accumulated phase of the photon travelling from source $A$ ($B$) to the $n$-th detector.
While the intensity is
\begin{equation}
I_n = |a_n|^2 = I_A + I_B + 2 \mathbf{Re} \alpha \beta^* e^{i\phi_n},
\label{In} 
\end{equation}
where $I_A=|\alpha|^2$, $I_B=|\beta|^2$, and $\phi_n=\phi_{An}-\phi_{Bn}$.
If the relative phase of $\alpha$ and $\beta$ is fixed, Eq.~(\ref{In}) predicts that interference fringes will be observed.
If there is no correlation between the phases of $\alpha$ and $\beta$, there are no interference fringes
because
\begin{equation}
\langle I_n\rangle = \langle I_A\rangle + \langle I_B\rangle. 
\end{equation}
On the other hand, the product of the intensities is given by
\begin{eqnarray}
I_nI_m = |a_na_m|^2 = |\alpha^2 e^{i(\phi_{An}+\phi_{Am})}+\beta^2 e^{i(\phi_{Bn}+\phi_{Bm})} \cr 
 + \alpha\beta(e^{i(\phi_{An}+\phi_{Bm})}+e^{i(\phi_{Am}+\phi_{Bn})}) |^2,
\end{eqnarray}
and after averaging over the uncorrelated phases of $\alpha$ and $\beta$, we find
\begin{eqnarray}
G_{nm}^{(2)} &=& \langle I_nI_m\rangle = \langle I_A I_A\rangle + \langle I_B I_B\rangle +\langle I_AI_B\rangle |e^{i(\phi_{An}+\phi_{Bm})}+e^{i(\phi_{Am}+\phi_{Bn})} |^2 \cr
&=& \langle I_A^2\rangle + \langle I_B^2\rangle + 2\langle I_AI_B\rangle (1+\cos\phi_{nm})
\label{InIm}
\end{eqnarray}
where $\phi_{nm}=\phi_n-\phi_m$ and $n,m=1,2,3$. 
According to Eq.~(\ref{InIm}) the intensity-intensity correlation
will exhibit interference fringes, a manifestation of the so-called Hanbury Brown-Twiss effect.
It is convenient to introduce the normalized, dimensionless, correlation by
\begin{equation}
g_{nm}^{(2)} \equiv \frac{G_{nm}^{(2)}}{\langle I_n\rangle\langle I_m\rangle},
\end{equation}
where $\langle I_n\rangle=\langle I_m\rangle=\langle I_A\rangle+\langle I_B\rangle$.
Assuming that the sources $A$ and $B$ have the same statistics and the same average intensities,
we have $I_A=I_B$ and obtain
\begin{equation}
g_{nm}^{(2)} =g^{(2)} \left( 1 +\frac{1}{2}\cos\phi_{nm}\right),
\label{eqhbt2d}
\end{equation}
where $g^{(2)}=\langle I_A^2\rangle/\langle I_A\rangle^2$ is the second-order normalized intensity autocorrelation function.
Similarly, we consider the averages of the product of three intensities given by
\begin{equation}
G_{123}^{(3)} = \langle I_1I_2I_3 \rangle
=\langle I_A^3 \rangle + \langle I_B^3 \rangle + 
[\langle I_A^2 \rangle\langle I_B \rangle+ \langle I_B^2 \rangle\langle I_A\rangle]
[3 + 2(\cos\phi_{12}+\cos\phi_{23}+\cos\phi_{13})],
\end{equation}
and, assuming $I_A=I_B$ as before, we have
\begin{equation}
g_{123}^{(3)} \equiv  \frac{G_{123}^{(3)}}{\langle I_1\rangle\langle I_2\rangle\langle I_3\rangle}
=\frac{g^{(3)}}{4}+\frac{g^{(2)}}{2}\left( \frac{3}{2}+\cos\phi_{12}+\cos\phi_{23}+\cos\phi_{13}\right),
\label{eqhbt3d}
\end{equation}
where $g^{(3)}=\langle I_A^3\rangle/\langle I_A\rangle^3$ is the third-order normalized intensity autocorrelation function.
In this paper, we consider the case of coherent light only. Then we have $g^{(3)}=g^{(2)}=1$.

\section{Event-by-event simulation}

We first discuss the general aspects of our event-by-event, particle-only simulation approach.
This approach is unconventional in that it does not require
knowledge of the wave amplitudes obtained by first solving the wave mechanical problem
nor do we first calculate the quantum potential (which
requires the solution of the Schr{\"o}dinger equation) and
then compute the Bohm trajectories of the particles.
Instead, the detector clicks are generated event-by-event by locally causal,
adaptive, classical dynamical systems.
Our approach employs algorithms, that is we define processes, that contain
a detailed specification of each individual event
which cannot be derived from a wave theory such as QT.

The simulation algorithms that we construct describe processes that are
most easily formulated in terms of events, messages,
and units that process these events and messages.
In a pictorial description, the photon is regarded as a messenger,
carrying a message that represents its time-of-flight.
In this pictorial description, we may speak of ``photons'' generating
the detection events. However, these so-called photons, as we will call them in the following,
are elements of a model or theory for the real laboratory experiment only.
The only experimental facts are the settings of the various apparatuses and the detection events.

The processing units mimic the role of the optical components in the experiment.
A network of processing units represents the complete experimental setup.
The standard processing units consist of an input stage,
a transformation stage and an output stage.
The input (output) stage may have several channels
at (through) which messengers arrive (leave).
Other processing units are simpler in the sense that the input stage
is not necessary for the proper functioning of the device.
A message is represented by a set of numbers, conventionally represented by a vector.
As a messenger arrives at an input channel of a processing unit,
the input stage updates its internal state, represented by a vector, and
sends the message together with its internal state
to the transformation stage that implements the operation of the particular device.
Then, a new message is sent to the output stage
which selects the output channel through which the messenger will leave the unit.
At any given time, there is only one messenger being routed through the whole network.
There is no direct communication between the messengers nor is there any communication
between the processing units other than through the messengers.
We view the simulation as a message-processing and message-passing process:
It routes messengers, representing the photons, through a network of message-processing units,
representing the optical components in the laboratory experiment.
From this general description, it should already be clear that the process that
is generated by the collective of classical dynamical systems is locally
causal in Einstein's sense.

\subsection{Simulation model}

The network of processing units represents the whole experimental setup.
For the present purposes, that is the demonstration that the HBT effect can
be explained by a particle-only model, it is sufficient to simulate the bottom part of Fig.~\ref{hbt2}.
All the components, photons, beam splitters and photon detectors, have corresponding parts in our event-based simulation.
As all the components are already presented in our previous
work~\cite{RAED05d,RAED05b,RAED05c,MICH05,RAED06c,RAED07a,RAED07b,RAED07c,ZHAO07b,ZHAO08,ZHAO08b,JIN09a,JIN09c},
for completeness, we only give a brief description of each of the components of the simulation setup.
 
\subsubsection{Messenger}
We view each photon as a messenger.
Each messenger has its own internal clock, the hand of which rotates with frequency $f$.
When the messenger is created, the time of the clock is set to zero. 
As the messenger travels from one position in space to another, the clock encodes the time of flight $t$ modulo the period $1/f$.
The message, the position of the clock's hand, is most conveniently represented by a two-dimensional unit vector
${\bf{e}}_l=(e_{0,l},e_{1,l})=(\cos \psi_l, \sin \psi_l)$, where $\psi_l=2\pi f t_l$ and the subscript $l>0$ labels the successive messages.
The messenger travels with the speed of light $c$.
In this paper, we do not need to specify the fixed frequency $f$ and to specify a message,
we use the angle $\psi_l$ instead of the time-of-flight $t_l$.

\subsubsection{Beam splitter}

\begin{figure}[t]
\begin{center}
\includegraphics[width=12.0cm]{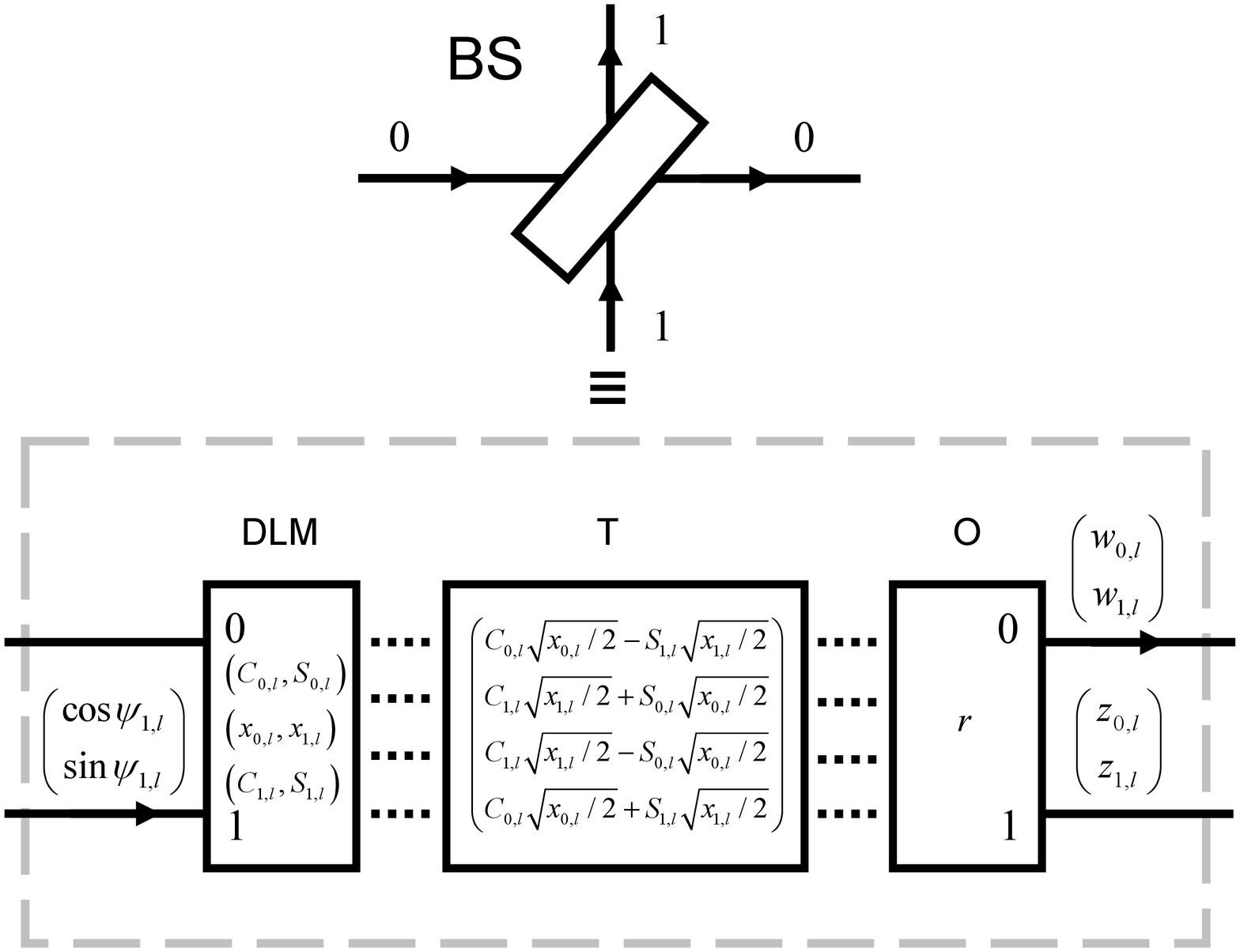}
\caption{Diagram of a DLM-based processing unit that performs an event-based simulation of a beam splitter (BS). 
The processing unit consists of three stages: An input stage (DLM), a transformation stage (T) and an output stage (O).
The solid lines represent the input and output channels of the BS. The dotted lines indicate the data flow within the BS.
}
\label{beamsplitter}
\end{center}
\end{figure}

The structure of the processing unit for a beam splitter (BS) is shown in Fig.~\ref{beamsplitter}.
The unit has two input and two output channels labeled by $k=0,1$ and consists of an input stage (DLM)
a transformation stage (T), and an output stage (O).

The input stage receives a message on either input channel $0$ or $1$, never on both channels simultaneously.
The input events are represented by the vectors ${\bf v}_l=(1,0)$ or ${\bf v}_l=(0,1)$ if the $l$th
event occurred on channel 0 or 1, respectively
and are processed by a simple deterministic learning machine (DLM)~\cite{RAED05d,RAED05b,RAED05c,MICH05,ZHAO08b}.
The DLM has two internal registers ${\bf Y}_{k,l}=(C_{k,l},S_{k,l})$ and one internal vector ${\bf x}_{l}=( x_{0,l},x_{1,l}) $,
where $x_{0,l}+x_{1,l}=1$ and $x_{k,l}\geq 0$ for $k=0,1$ and all $l \geq 0$.
Upon receiving the $l$th input event, the DLM performs the following steps:
It stores the input message ${\bf e}_{k,l}=(\cos\psi_{k,l},\sin\psi_{k,l})$ in
its internal register ${\bf Y}_{k,l}=(C_{k,l},S_{k,l}) $.
Then, it updates its internal vector according to the rule
\begin{equation}
{\bf x}_{l}=\gamma {\bf x}_{l-1}+( 1-\gamma ) {\bf v}_l,
\label{eq_x}
\end{equation}
where $0<\gamma <1$.
A detailed analysis of the update rule Eq.~(\ref{eq_x}) can be found in Ref.~\cite{JIN09a}.

The transformation stage accepts the messages from the input stage, and transforms them into a new four-dimensional vector
\begin{equation}
{\bf T}=\frac{1}{\sqrt{2}}
\left(
\begin{array}{c}
C_{0,l}\sqrt{x_{0,l}} - S_{1,l}\sqrt{x_{1,l}} \\
C_{1,l}\sqrt{x_{1,l}} + S_{0,l}\sqrt{x_{0,l}} \\
C_{1,l}\sqrt{x_{1,l}} - S_{0,l}\sqrt{x_{0,l}} \\
C_{0,l}\sqrt{x_{0,l}} + S_{1,l}\sqrt{x_{1,l}} \\
\end{array}
\right).
\label{T_vector}
\end{equation}%

The output stage sends out a messenger (representing a photon) carrying the message
\begin{equation}
{\bf w}=\left(
\begin{array}{c}
w_{0,l} \\
w_{1,l}
\end{array}%
\right) ,
\end{equation}%
where
\begin{eqnarray}
w_{0,l}&=&\left.\left(C_{0,l}\sqrt{x_{0,l}/2} - S_{1,l}\sqrt{x_{1,l}/2}\right)\right/s_l,\cr 
w_{1,l}&=&\left.\left(C_{1,l}\sqrt{x_{1,l}/2} + S_{0,l}\sqrt{x_{0,l}/2}\right)\right/s_l,\cr 
s_{l}&=&\sqrt{w_{0,l}^{2}+w_{1,l}^{2}}. 
\end{eqnarray}
through output channel 0 if $s_{l}^2>r$ where
$0<r<1$ is a uniform pseudo-random number.
Otherwise, if $s_{l}^2 \le r$, the output stage sends
through output channel 1 the message
\begin{equation}
{\bf z}=\left(
\begin{array}{c}
z_{0,l} \\
z_{1,l} 
\end{array}
\right),
\end{equation}
where
\begin{eqnarray}
z_{0,l}&=&\left.\left(C_{1,l}\sqrt{x_{1,l}/2} - S_{0,l}\sqrt{x_{0,l}/2}\right)\right/t_l,\cr 
z_{1,l}&=&\left.\left(C_{0,l}\sqrt{x_{0,l}/2} + S_{1,l}\sqrt{x_{1,l}/2}\right)\right/t_l,\cr 
t_{l}&=&\sqrt{z_{0,l}^{2}+z_{1,l}^{2}}.
\end{eqnarray}
We use pseudo-random numbers to mimic the apparent unpredictability
of the experimental data only: The use of pseudo-random numbers to select the output channel
is not essential~\cite{RAED05b}. 
Note that in our simulation model there is no need to introduce
the (quantum theoretical) concept of a vacuum field, a requirement in the quantum optical description of a BS.

\subsubsection{Photon detector}

\begin{figure}[t]
\begin{center}
\includegraphics[width=12cm]{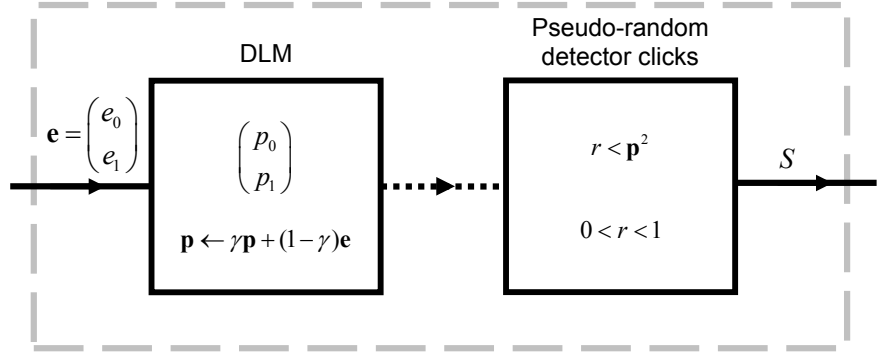}
\caption{%
Diagram of the event-based detector model defined by Eqs.~(\ref{ruleofLM}) and (\ref{thresholdofLM}).
The dotted line indicates the data flow within the processing unit.
}%
\label{detector}
\end{center}
\end{figure}

A schematic diagram of the unit that functions as a single-photon 
detector is shown in Fig.~\ref{detector}~\cite{JIN09a}.
The first stage consists of a DLM 
that receives on its input channel the $l$th message represented by
the two-dimensional vector ${\mathbf e}_l=(\cos \psi_l,\sin\psi_l)$. 
In this paper, we use the simplest DLM containing a single
two-dimensional internal vector with Euclidean norm less or equal than one.

We write ${\mathbf p}_l = (p_{0,l},p_{1,l})$ to denote
the value of this vector after the $l$th message has been received.
Upon receipt of the $l$th message the internal vector is updated according to the rule
\begin{equation}
	{\mathbf p}_l = \gamma {\mathbf p}_{l-1} + (1-\gamma) {\mathbf e}_l,
	\label{ruleofLM}
\end{equation}
where $0<\gamma<1$ and $l>0$. 
If $\gamma\not=0$, a machine  that operates
according to the update rule Eq.~(\ref{ruleofLM}) has memory to store 
an amount of information that is equivalent to the information carried by
a single mesasage only.
Obviously, the rule Eq.~(\ref{ruleofLM}) is the same as that used for the BS (see Eq.~(\ref{eq_x}))
but the input data is different.

The second stage of the detector (see Fig.~\ref{detector}) uses the information
stored in the internal vector to decide whether or not to generate a click.
As a highly simplified model for the bistable character of the real photodetector or
photographic plate, we let the machine generate a binary output signal $S_k$ using the threshold function
\begin{equation}
	S_l = \Theta({\mathbf p}^2_l-r_l),
	\label{thresholdofLM}
\end{equation}
where $\Theta(.)$ is the unit step function and $0\leq r_l <1$ is a uniform pseudo-random number.
Note that in contrast to experiment, in a simulation, we could register both the $S_l=0$ and $S_l=1$ events 
such that the number of input messages equals the sum of the $S_l=0$ and $S_l=1$ detection events.
Since in experiment it cannot be known whether a photon has gone undetected, we discard the information about
the $S_l=0$ detection events in our future analysis.
The total detector count is defined as
\begin{equation}
	N=\sum^{l}_{j=1}S_j,
	\label{N_counts}
\end{equation}
where $l$ is the number of messages received. Thus, $N$ counts the number of one's generated by the machine.

\subsubsection{Experiment}

The processing units that simulate the optical components are connected in such a way 
that the network corresponds to the experimental set up in the laboratory.
As explained earlier, it is sufficient to consider the bottom part of Fig.~\ref{hbt2}.

\section{Simulation results}

Following Ref.~\cite{AGAF08}, the phase of the coherent photons emitted by the source
is ``randomized'' by letting the light pass through an EOM,
the voltage of which is switched with a frequency of $50$ Hz.
To mimic this in the simulation, we send $N_{interval}$ messengers with some fixed but randomly chosen phase,
then another $N_{interval}$ messengers with another fixed but randomly chosen phase, and so on.
In practice, we use $N_{interval}=2500$.
The messengers (photons) are sent through either channel $A$ or $B$, one at a time
and are either transmitted or reflected by the beam splitters.
Before hitting a detector, the messenger experiences a time delay corresponding to $\phi_{An}$ or $\phi_{Bn}$ ($n=1,2,3$).
The detector processes the message carried by the messenger and decides whether or not to produce a click.

We consider three different experiments.
In case 1, we remove all BSs in Fig.~\ref{hbt2}(bottom) and study the signal produced by detector $D_1$.
Then, in case 2, we remove BS2, that is we consider the HBT experiment with two detectors, as indicated
by the dashed-dotted line in Fig.~\ref{hbt2}(bottom).
Finally, in case 3, we study the full three-photon correlation experiment, see Fig.~\ref{hbt2}(bottom).
In cases 2 (3), the intensity-intensity correlations are calculated
by counting coincidences of two (three) messengers, meaning
that the arrival times of the two (three) messagers are within a time window $W$, to be discussed in Section~4.4.
All simulations have been carried out with $\gamma=0.99$.

\subsection{Case 1: One detector}

\begin{figure}[t]
\begin{center}
\includegraphics[width=12cm]{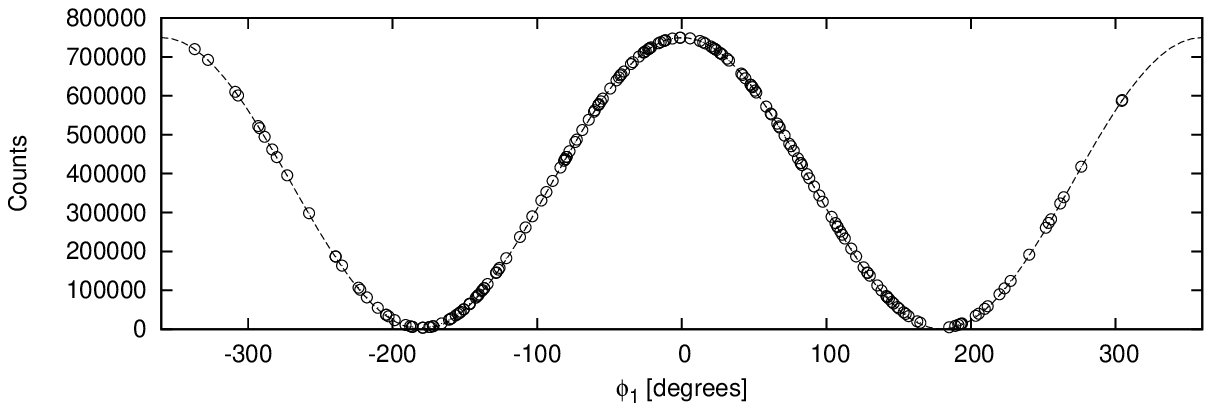}
\includegraphics[width=12cm]{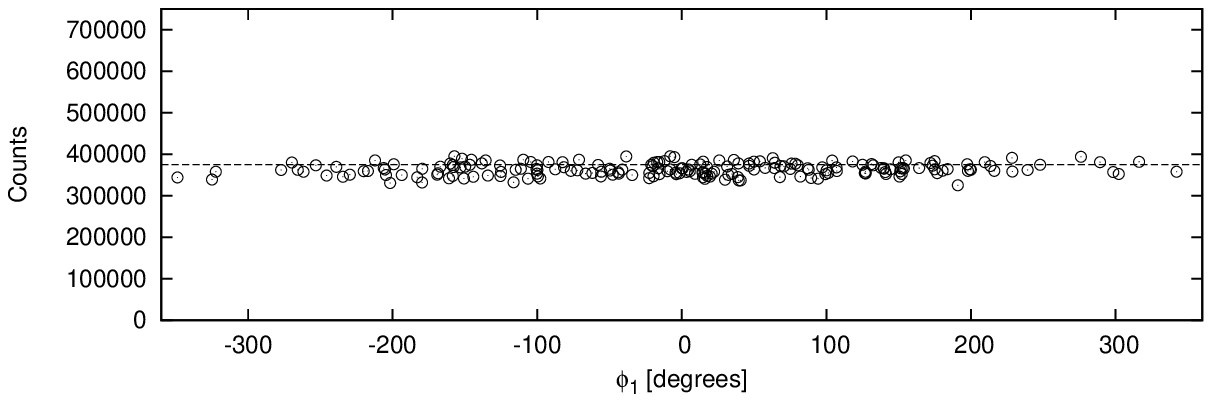}
\caption{Case 1: All BSs in Fig.~\ref{hbt2}(bottom) removed.
Simulation results for the detector counts as a function of $\phi_1=\phi_{A1}-\phi_{B1}$.
Top: The differences between 
the time-of-flights of the messengers entering channel $A$ and
the time-of-flights of the messengers entering channel $B$ 
is constant.
Bottom: The differences between 
the time-of-flights of the messengers entering channel $A$ and
the time-of-flights of the messengers entering channel $B$ are random.
Circles: simulation data; Dashed line: Wave theory solution Eq.~(\ref{In}) averaged over $\phi_1$.
}
\label{rds}
\end{center}
\end{figure}

Let us first demonstrate how the event-based model of the detector works~\cite{JIN09a}.
The messengers, randomly entering through channels $A$ or $B$, are sent directly to the 
time-delay units that change the angle, representing the time-of-flight, 
by $\phi_{A1}$ or $\phi_{B1}$, respectively.
The messengers are then processed by detector $D_1$.
We perform two different sets of simulations.
First, we keep the differences between 
the time-of-flights of the messengers entering channel $A$ and
the time-of-flights of the messengers entering channel $B$ 
constant.
In this case, according to wave theory, we expect to see clear interference fringes.
Second, the differences between 
the time-of-flights of the messengers entering channel $A$ and
the time-of-flights of the messengers entering channel $B$ are taken to be random.
Then, according to wave theory, there should be no sign of interference effects.
As shown in Fig.~\ref{rds}, our particle-only simulation approach reproduces both features
and the results are in very good agreement with the wave theoretical results (see Eq.~(\ref{In})).

\subsection{Case 2: Hanbury Brown-Twiss experiment}

\begin{figure}[t]
\begin{center}
\includegraphics[width=12cm]{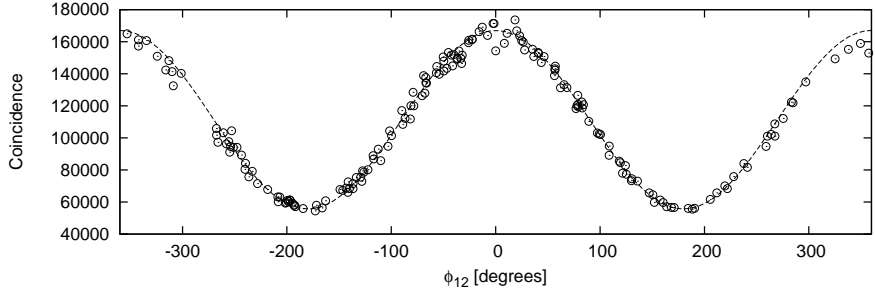}
\caption{Case 2: BS2 in Fig.~\ref{hbt2}(bottom) removed.
Simulation results of the two-particle coincidence counts as a function of
$\phi_{12}$ where $\phi_{12}=\phi_{1}-\phi_{2}$, and $\phi_n=\phi_{An}-\phi_{Bn}$ ($n=1,2$).
The time-of-flights of the messengers entering channel $A$ and
the time-of-flights of the messengers entering channel $B$ are taken to be random.
Circles: simulation data; Dashed line: wave theory solution Eq.~(\ref{eqhbt2d}).
}
\label{rhbt2}
\end{center}
\end{figure}

We consider the HBT experiment with two detectors, that is
we remove BS2 from the diagram in Fig.~\ref{hbt2}(bottom).
Messengers enter the apparatus through channel $A$ or $B$, one by one.
The time-of-flights of the messengers entering channel $A$ and
the time-of-flights of the messengers entering channel $B$ are taken to be random
hence, as shown in Fig.~\ref{rds}(bottom) there is no first-order interference.
When passing a BS, the message changes according to the rules explained in Section~3.2.1.
Then, before entering the detector, the message is changed once more
by $\phi_{An}$ or $\phi_{Bn}$ ($n=1,2$), depending on which path the messenger took.
If the two detectors fire with the time window $W$ (see Section~4.4),
we increase the number of coincidences.
The simulation data shown in Fig.~\ref{rhbt2} confirm that this
procedure reproduces the results of wave theory, see Eq.~(\ref{eqhbt2d}).

\subsection{Case 3: Three-particle intensity-intensity correlation}

\begin{figure}[t]
\begin{center}
\includegraphics[width=12cm]{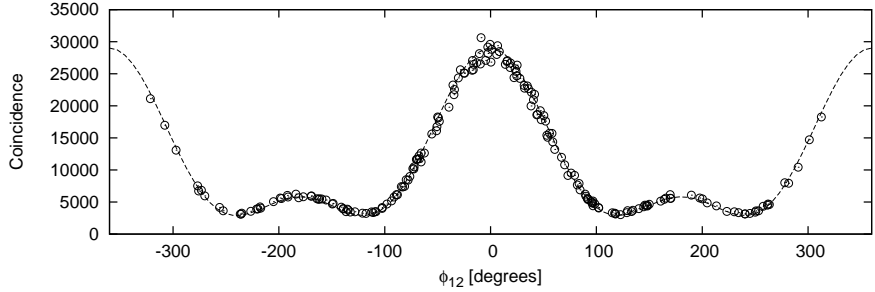}
\caption{Case 3: Three particle correlation experiment (see Fig.~\ref{hbt2}(bottom)).
Simulation results of the three-particle coincidence counts as a function of
$\phi_{12}$ where $\phi_{12}=\phi_{1}-\phi_{2}$, and $\phi_n=\phi_{An}-\phi_{Bn}$ ($n=1,2,3$).
We only show data for the case $\phi_{A2}=\phi_{B2}=0$, $\phi_{A1}=\phi_{B3}$, $\phi_{B1}=\phi_{A3}$
where $\phi_{A1}$ and $\phi_{B1}$ are chosen randomly.
The time-of-flights of the messengers entering channel $A$ and 
the time-of-flights of the messengers entering channel $B$ are taken to be random.
Circles: simulation data; Dashed line: wave theory solution Eq.~(\ref{eqhbt3d}).
}

\label{rhbt3}
\end{center}
\end{figure}

Finally, we consider the full correlation experiment Fig.~\ref{hbt2}(bottom) with three detectors.
The simulation procedure is the same as in case 2, except that we count coincidences of 
clicks of three different detectors.
Also in this case, the simulation data shown in Fig.~\ref{rhbt3} confirm that this
procedure reproduces the results of wave theory, see Eq.~(\ref{eqhbt3d}).

\subsection{Discussion}

Our simulation model is based on a particle picture and makes no reference to concepts or results from wave theory.
In contrast to the conventional quantum theoretical explanation in terms of the
wave-particle nature of photons, our simulation approach requires a particle picture of photons only.  
During the event-by-event simulation we always have full which-way information 
of the photons (messengers) since we can always track them.
Nevertheless, depending on the settings of the optical apparatuses, intensity-intensity interference is observed.
Although the appearance of an interference pattern is commonly considered
to be characteristic of a wave, we have demonstrated that, as in experiment, 
it can also appear as a result of a collection of particles that interact
with the various optically active devices such as beam splitters and detectors.
In this paper, we considered the case that is equivalent to a light source that produces photons in a coherent state only. 
The case of a thermal light source will be considered in future work.

In real experiments, and also in our simulation approach, it is necessary to specify the procedure
by which we count coincidences of detection events. 
For the experiments at hand, one introduces a time window $W$
and one defines as a two (three) particle coincidence, two (three) detection events with the time difference(s) 
are smaller than $W$.
As discussed extensively in our work on the simulation of Einstein-Podolsky-Rosen-Bohm (EPRB) experiments~\cite{RAED07c}, 
the choice of the time window $W$ is of crucial importance, both in the simulation and in real experiments~\cite{WEIH98},
to obtain the correlation of a quantum system in the singlet state.
In general, only when $W\rightarrow 0$, experiment and simulation can
reproduce the correlation of a quantum system in the singlet state~\cite{RAED07c}.
For large enough $W$, the relation to a quantum system in the singlet state is lost.
In this paper, we have chosen $W$ sufficiently large and generated groups of two (three) messengers 
such that if the two (three) detectors fire, this constitutes a coincidence of two (three) particles.
In other words, the time delays are only used by the detector but are ignored in determining coincidences.
In this sense, the simulation results presented in this paper pertain
to classical light and are therefore in excellent agreement with classical wave theory.
To study the quantum aspects of two- and three-particle correlations
the time delays should be used to also determine the coincidences,
as in our EPRB simulations~\cite{RAED07c}.
We leave this very interesting topic for future research.

\section{Conclusion}

We have demonstrated that our classical, locally causal, particle-like simulation
approach reproduces the results of the Hanbury Brown-Twiss effect.
Our event-based simulation model, a classical, locally causal, adaptive dynamical system,
reproduces the results of wave theory without making reference to the solution
of a wave equation and provides a simple, particle-based mental picture for what each
individual photon experiences as it travels from the source to the detector.
Our simulation algorithm demonstrates that the wave-particle duality is not the only way to describe the nature
of a phtone but that there is another way that only needs the particle nature, 
satisfies Einstein's local causality and does not defy the common sense.
Finally, we would like to emphasize that the algorithms used to simulate
the optical components in this paper have not been designed to simulate the HBT-type experiments.
The algorithms have been taken, without modification, from our earlier work on very different 
quantum optics experiments~\cite{RAED05d,RAED05b,RAED05c,MICH05,RAED06c,RAED07a,RAED07b,RAED07c,ZHAO07b,ZHAO08,ZHAO08b,JIN09a,JIN09a,JIN09c}.
In this sense, it seems that our approach has predictive power: The algorithms can be reused to simulate very different experiments
than those for which they were originally developed.

\bibliographystyle{unsrt}
\bibliography{../epr}

\end{document}